\documentclass[prb,twocolumn,showpacs,superscripaddress]{revtex4}
\usepackage{epsfig}
                                                                                
\begin{document}
                                                                                
\title{
Cooper-pair transport through a Hubbard chain sandwiched between
    two superconductors: Density matrix renormalization group
    calculations
}

\author{Adrian E.~Feiguin} 
                                                                                
\affiliation{Microsoft Project Q\\
The University of California at Santa Barbara, Santa Barbara, CA 93106}
\email{afeiguin@microsoft.com}

\author{Steven R.~White}
\affiliation{Department of Physics and Astronomy\\
University of California, Irvine, CA 92697}
\email{srwhite@uci.edu}
                             
\author{D. J. Scalapino}

\affiliation{Department of Physics\\ 
University of California, Santa Barbara, CA 93106-9530}
\email{djs@vulcan2.physics.ucsb.edu}
                                                   
\date{\today}
                                                                                
\begin{abstract}
We present a numerical approach to study the coherent transport of Cooper pairs
through a Hubbard chain, and study the role of the contacts in achieving 
perfect Andreev reflection. 
We calculate the pair transport using the Density Matrix 
Renormalization Group by measuring the response
of the system to quantum pair fields with complex phases on the two
ends of an open system. This approach gives an effective superfluid 
weight which is in close agreement with the 
Bethe Ansatz results for the superfluid weight for closed Hubbard rings. 
\end{abstract}
                                                                                
\maketitle

\section{Introduction}

Pair transport through interacting one-dimensional
systems sandwiched between two superconductors has been the focus of much 
attention recently. These systems not only exhibit interesting 
physical phenomena, such as Andreev reflection and pair transport, but 
also may be incorporated into novel nanoelectronic devices.
In particular, there have been various proposals for the creation and
transport of entangled pairs for quantum communication. Therefore, it is 
useful to have well-controlled numerical tools for analyzing phenomena 
associated with pair transport through a 
superconducting-Hubbard-superconducting (SHS) system. 

In this paper we present the results of numerical studies 
of a one-dimensional Hubbard chain sandwiched between two 
superconducting contacts. We examine the effects of the boundary contacts 
on 
the injection and transport of pairs through the system.
We then introduce the idea of an extended contact between the 
superconductors and the intervening Hubbard chain which provides for 
improved pair transmission into and out of the Hubbard chain.  Using this 
we present a new numerical method for determining the effective superfluid
weight $\tilde{D}$ of a Hubbard chain.

\section{The Contact Model}

There have been various analytic studies of a one-dimensional Luttinger 
liquid sandwiched between two
superconductors \cite{Maslov 1996, Fazio 1996, Takane 1997}. Here, we make 
use of a comprehensive analysis recently reported by Affleck {\it et.~al} 
\cite{Affleck 2000}.  These authors integrated out the electron 
degrees of freedom of the superconducting leads, replacing them with 
effective boundary conditions for the Luttinger liquid. In this framework, 
the effective Hamiltonian for an SHS system can be written as
\begin{equation}
H=H_0+H_1
\label{one}
\end{equation}
where $H_0$ corresponds to the Hubbard chain
\begin{equation}
H_0=-t \sum\limits^L_{i=1\atop{\sigma}} \left(c^\dagger_{i\sigma} c_{i+1\sigma}+h.c.\right)
+ U \sum\limits^L_{i=1} n_{i\uparrow} n_{i\downarrow} - \mu \sum\limits^L_{i=1} n_i
\label{two}
\end{equation}
and $H_1$ incorporates the effects of the two superconducting leads
\begin{eqnarray}
H_1 & = & \Delta_L \left(e^{i\phi_L} c^\dagger_{1\uparrow} c^\dagger_{1\downarrow} +
h.c.\right) + \Delta_R \left(e^{i\phi_R} c^\dagger_{L\uparrow} c^\dagger_{L\downarrow} +
h.c.\right)\nonumber\\
& + & V_1(n_{1\uparrow} + n_{1\downarrow}) + V_L (n_{L\uparrow} + n_{L\downarrow})\, .
\label{three}
\end{eqnarray}
Here, $c^\dagger_{\ell\uparrow}$ creates an electron of spin up on the 
$\ell^{\rm th}$ site. The hopping parameter of the Hubbard chain is $t$, 
$U$ is the onsite interaction energy, and $\mu$ is the usual chemical 
potential. As discussed in reference \cite{Affleck 2000}, the effect of the two
superconducting leads can be parametrized in terms of contact pairing 
strengths $\Delta_{(L,R)}$ and their phases $\phi_{(L, R)}$, along with 
end point scattering potentials $V_1$ and $V_L$. 

In the following we will be interested in the symmetric case in
which $\Delta_L=\Delta_R=\Delta$ and $V_1=V_L=V$. The 
first term in $H_1$ 
injects or removes pairs with different phases on both ends. In addition, 
there are  effective boundary scattering potentials $V_{(1, L)}$ which 
arise and play an important role in achieving
optimal pair transmission across the ends of the Hubbard chain. 
Integrating out the superconducting electron degrees-of-freedom can be 
seen as a natural thing to do when the
Fermi level lies well below the superconducting gaps in the bulk of the 
superconductors,
since the pair fields in the superconductors have well-defined average 
values and negligible fluctuations. In the Hubbard system, the value of the 
pair field has to be replaced by the fluctuating pair operator. A similar 
approach was used by Kozub \cite{Kozub 2003} to study Josephson 
transport through a Hubbard impurity center.  

In their paper, Affleck, {\it el.~al} \cite{Affleck 2000} calculate the 
Josephson current and the Andreev reflection probability. For the
non-interacting half-filled tight-binding chain, they find that the 
maximum transmission probability 
is 1 (perfect Andreev reflection) and it occurs when $\Delta=t$ and 
$V=0$. In this case, the Josephson current versus the phase difference 
$\phi=\phi_R-\phi_L$ between the ends has Ishii's
sawtooth form \cite{Ishii 1970}. For smaller values of $V$, the sawtooth 
is smoothed out and starts resembling the Josephson sine shape 
corresponding to a small Andreev reflection probability (See also, 
Ref.~\cite{Caux 2002}). Away from half-filling, Affleck {\it et.~al} found 
that in order to achieve perfect Andreev reflection, both the contact 
pairing strength $\Delta$ and the boundary scattering potential $V$ needed 
to be tuned to particular values. For the non-interacting case, these 
values are:
\begin{equation}
V = \frac{\mu}{2};\ \Delta =t\sqrt{1-\frac{\mu^2}{4t^2}}\, .
\label{four}
\end{equation} 

In order to treat the interacting case, these authors employed 
bosonization and renormalization group methods. For negative values of 
$U$, they showed that the contact Hamiltonian renormalizes to the perfect 
Andreev reflection fixed point. Thus, even when the parameters of the 
contact were not fine-tuned for perfect Andreev reflection, one recovers 
the sawtooth form for the Josephson current versus the phase difference as 
the length $L$ of the Luttinger liquid increases. However, for
positive $U$, they found that the contact Hamiltonian flows away from the 
Andreev fixed point. In this case, as $L$ increases, the effective 
coupling of the superconductor to the
Luttinger liquid renormalizes to zero. For a finite value of $L$ and $U>0$, the
 coupling is weak and one 
finds the usual $J_1\sin\phi$ Josephson relation.  As $L$ increases, $J_1$
rapidly decreases and the transport of pairs through the chain vanishes in the
$L\to\infty$ limit.

\section{The Effective Superfluid Weight $\tilde{D}$}

In the following numerical study, we will be interested in determining an 
effective superfluid weight $\tilde{D}$. 
If the pair phase varies linearly across a Hubbard chain of length $L$, 
then there will be a uniform Josephson current, and we will define 
$\tilde{D}(L)$ 
by
\begin{equation}
j=\tilde{D} \frac{\phi_0}{L}
\label{five}
\end{equation}
with $\phi_0$ the phase difference across the Hubbard chain. The effective 
superfluid weight $\tilde{D}$ is then given by $\tilde{D}(L)$ as $L \rightarrow 
\infty$. Here, we have set $e=\hbar=1$.
The problem of determining $\tilde{D}(L)$ is to create a linear phase change 
$\phi_0/L$ across
the Hubbard chain and then to measure $j$. The latter is straightforward since 
\begin{equation}
j_i=-i[H, n_i]
\label{six}
\end{equation}
so that for $i\not= 1$ or $L$,
\begin{equation}
j_i=-it\sum\limits_\sigma \left(c^\dagger_{i\sigma} c_{i+1\sigma}-c^\dagger_{i+1\sigma}
c_{i\sigma}\right)\, .
\label{seven}
\end{equation}
At the boundary, when $i=1$ (or $L$) we have to consider the boundary terms and add an extra
current operator
\begin{equation}
j_1^{\prime}=-iV(\exp(i\phi)c_{1\uparrow}^{\dagger}c_{1\downarrow}^{\dagger}
- \exp(-i\phi) c_{1\downarrow}c_{1\uparrow})\, .
\label{eight}
\end{equation}
with a similar term for the right hand $i=L$ boundary.
The current density is independent of the position, and any
of these expressions can be used with these end corrections to calculate $j$. 

The measurement of $j$ is straightforward within the DMRG 
method\cite{White 1992, White 1993}. However, it is also
necessary to establish a uniform phase gradient.  As noted in the previous 
section, for a finite length $L$ of the Hubbard chain, this can require 
tuning of the contact boundary pairing strength and the boundary 
scattering potential. Fortunately, for negative values of
$U$, the contact interaction renormalizes to the perfect Andreev 
reflection fixed point as the length of the chain increases.  However, 
when the finite system is doped away from half-filling, there are two 
parameters to tune and achieving a match such that the phase
gradient over the length $L$ is uniform becomes more difficult.  For this 
reason, we have developed an approach based upon extended contact 
interaction which will be discussed at the end of the next section.

\section{Results}

We use the Lanczos method for a system of size $L=8$ and DMRG for larger systems. The DMRG method
is the standard finite-size algorithm, except for the use of complex numbers due to the
arbitrary Josephson phases, and a special treatment of quantum numbers. The non-particle 
conserving boundary conditions mean that the total number of fermions cannot be used as a
conserved quantum number. However, one can still utilize the number of fermions modulo
2. 
This modulo-2 approach was first used in
\cite{threeleg}. Within this approach the local pair-field $\Delta$ can take on a definite
nonzero value. We have typically kept $m=200$ states per block for the results
presented, with a truncation error of about $10^{-8}$.

\subsection{Point Contacts}

\begin{centering}
\begin{figure}
\epsfig {file=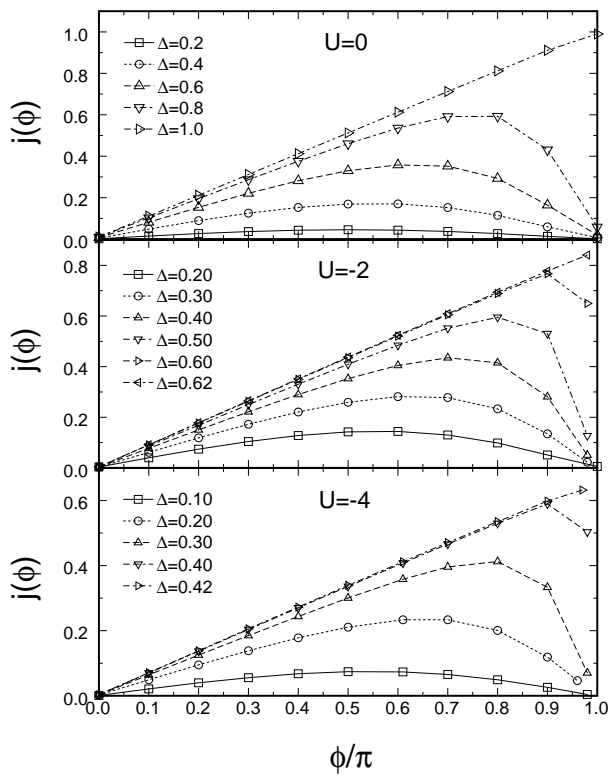,width=80mm}
\caption{Josephson current though a half-filled Hubbard chain with $L=8$
as a function of the phase $\phi$ and for
different values of the contact pairing $\Delta$, and Coulomb interaction 
$U$, in units where the hopping $t=1$.
}
\label{fig1}
\end{figure}
\end{centering}

In Fig.\ref{fig1} we show Lanczos results for the Josephson current versus 
$\phi=\phi_R-\phi_L$ through a half-filled Hubbard chain of $L=8$ sites.  
For this half-filled, particle-hole symmetric case, with $U\leq 0$, the required site 
potential $V_{1,L}=0$ and the contact pairing strength $\Delta$
can be adjusted to achieve perfect Andreev reflection. 
For the non-interacting case, this is obtained for $\Delta/t=1$ as shown in 
the top panel of Fig.\ref{fig1}. For negative values of $U$, it is 
necessary to fine-tune $\Delta$. When perfect Andreev reflection is 
achieved, $j_1(\phi)$ exhibits a sawtooth form with $j(\phi)=\tilde{D}(L)\phi/L$ 
for
$-\pi \leq \phi \leq \pi$. In this case, $\tilde{D}(L)$ can be directly 
determined 
from $j(\phi)$. For negative values 
of $U$, $\tilde{D}$ rapidly approaches its asymptotic value when $L\gg \pi 
t/|U|$, so that the important requirement for determining $\tilde{D}$ is to 
achieve perfect Andreev reflection at the ends.

\begin{centering}
\begin{figure}
\epsfig {file=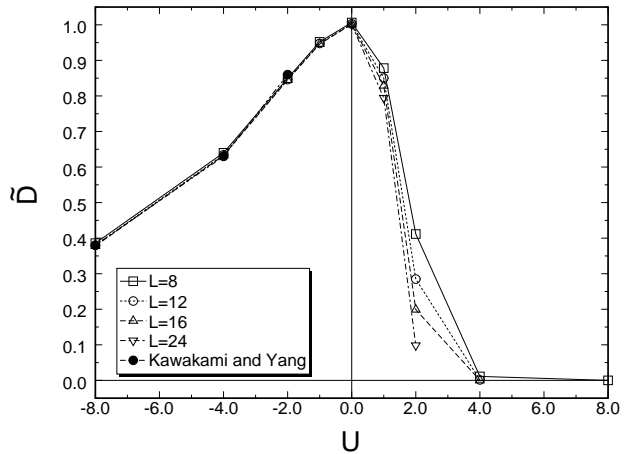,width=70mm,angle=-90}
\caption{Effective superfluid weight $\tilde{D}(L)$ of the half-filled Hubbard 
chain as a function of the Coulomb interaction $U$, for $\Delta=1$, $V=0$, and 
chains of various lengths.
We add for comparison the exact $L\to\infty$ Bethe Ansatz results from \cite{Kawakami 
and Yang} for $U<0$.
}
\label{fig2}
\end{figure}
\end{centering}

In Figure \ref{fig2}, we show Lanczos and DMRG results for the superfluid 
weight $\tilde{D}(L)$ of the half-filled chain for different values of the 
Coulomb interaction $U$. Here we have set
$V=0$ and taken $\Delta=1$. The renormalization to perfect Andreev 
reflection is rapid for $U<0$ and the resulting effective superfluid 
weight $\tilde{D}(L)$ varies little with $L$ giving a value in close 
agreement with the exact Bethe-Ansatz results for the superfluid weight of 
the infinite system, taken from \cite{Kawakami and Yang}.  
For $U>0$, the system renormalizes as $L$ increases to the non-superconducting
fixed point and the Josephson current is rapidly suppressed.

\begin{centering}
\begin{figure}
\epsfig {file=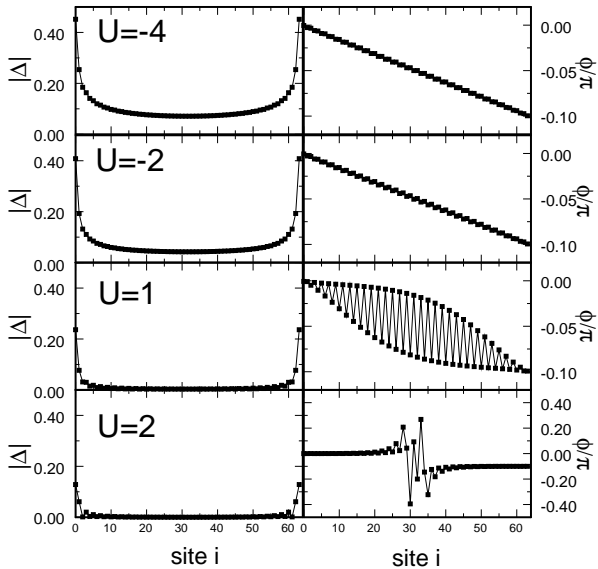,width=80mm}
\caption{Pair field amplitude and phase along the $L=64$ chain
for different values of $U$ at half-filling.
}
\label{fig3}
\end{figure}
\end{centering}

In Fig.\ref{fig3} we show the DMRG results for the pair field
amplitude along a chain at half-filling. The phase clearly varies
linearly for negative $U$, while for positive values the modulus
decays in a very short distance, a signature of the absence of 
superconductivity, and the phase order is disrupted by the small 
amount of noise in the DMRG.

For the 8-site chain, we have seen that for the half-filled, particle-hole 
symmetric case it is necessary to tune the contact pairing strength 
$\Delta$  in order to achieve perfect Andreev reflection. For the non 
half-filled case, for finite $L$, there are two contact coupling parameters,
$\Delta$ and $V$, that require tuning.  

\begin{centering}
\begin{figure}
\epsfig {file=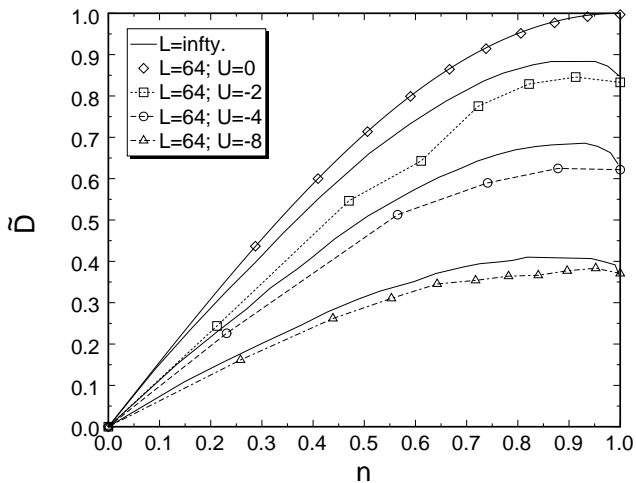,width=70mm,angle=-90}
\caption{$\tilde{D}$ as a 
function of the electron density for different values
of the Coulomb interaction $U$. For $U=0$, the boundary fields have been 
adjusted using Eqs.(\ref{four}) to achieve perfect Andreev reflection, 
while for $U<0$ we used $\Delta=1$, and $V_{1,L}=0$. 
We add for comparison the exact Bethe-Ansatz results from \cite{Kawakami 
and Yang} (solid lines).
}
\label{fig4}
\end{figure}
\end{centering}

In Fig.\ref{fig4} we show DMRG results for the superfluid
weight versus electron density and various values
of the Coulomb interaction. Here, 
$n$ is the electron density in the bulk of the chain, {\it i.e.} 
the center of the chain and far from the contacts). 
For comparison we show results for the superfluid weight $D_s$ for 
$L\to\infty$ obtained from 
Bethe Ansatz calculations \cite{Kawakami and Yang}. 
For $U=0$ we have adjusted the values of
$\Delta$ and $V$ for maximum transmitivity, Eq.~\ref{four}. 
For finite $U$ we have set $\Delta=1$ and $V_{1,L}=0$, {\it i.e.}
 they are not {\it optimized for perfect 
reflection}.

\begin{centering}
\begin{figure}
\epsfig {file=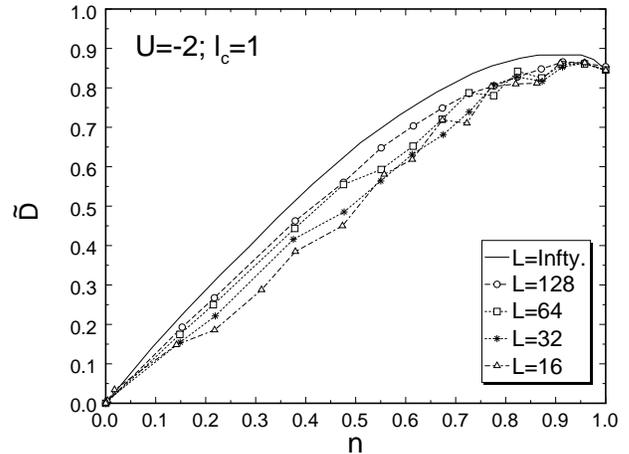,width=65mm,angle=-90}
\caption{Effective superfluid weigth $\tilde{D}$ of a Hubbard chain with 
$U=-2$, connected to point contacts, as a function
of density $n$, and for different lengths $L$. 
We add for comparison the Bethe Ansatz results in the thermodynamic limit 
(solid line).
}
\label{fig5}
\end{figure}
\end{centering}

In Fig.\ref{fig5} we show plots of $\tilde{D}$ versus $n$ for $U=-2$ and
chains of different lengths $L$. As $L$ increases, $\tilde{D}$ approaches
the exact result as the point contact boundary condition renormalizes to 
perfect Andreev reflection. However, to control this convergence it is in 
principle necessary to extrapolate the result to zero DMRG truncation error 
(large number of states $m$) and then take the infinite length limit \cite{extrap}.
Hence, we would expect these curves to be more accurate if we were to 
fine-tune the parameters. However, this task has proven to be difficult. 
In order to overcome the difficulties of fine 
tuning the parameters in the Hamiltonian for optimal transmitance, we 
have studied the effects of using extended smooth contacts at the 
boundaries.

\subsection{Extended contacts}

In the previous section we have discussed a Hubbard chain of finite length $L$
connected to superconductors through point contacts. We have seen that it
is necessary to tune the pairing strength $\Delta$ and the boundary scattering 
potential $V$ in order to obtain a linear phase change along the chain.
In this section, we explore the effects of extended contacts as an alternative way to eliminate
the normal contact reflection. This technique is inspired by the smooth boundary conditions 
approach\cite{Vekic 1993}.
Here, we have applied the pair field end terms over a length $\ell_c$ on 
the end of each chain, with the coefficient dropping smoothly to zero as 
the distance from the end approaches $\ell_c$.
We have
\begin{eqnarray}
H_1 & = & \sum\limits^{\ell_c}_{\ell=1} \Delta(\ell) 
\left(e^{i\phi_L}
c^\dagger_{\ell\uparrow} c^\dagger_{\ell\downarrow} + h.c.\right) 
\nonumber \\
& + & \sum\limits^L_{\ell=L-\ell_{c+1}} \Delta(L-\ell)
\left(e^{i\phi_R} c^\dagger_{\ell\uparrow} c^\dagger_{\ell\downarrow} + h.c.\right)
\nonumber\\
& + & \sum\limits^{\ell_c}_{\ell=1} V(\ell) n_{\ell} + \sum\limits^L_{\ell=L-\ell_c+1} V(L-\ell+1)
n_{\ell}\, 
\label{nine}
\end{eqnarray}
where we take $\Delta(x)=\Delta (1+\cos(x \pi/ \ell_c))/2$\cite{smoothfun}.
In the following, we will set $V(x)=0$ and examine various widths $\ell_c$ 
of the contact.

In calculating the superfluid weight with the extended contacts one must 
utilize only the local properties in the center of the system.
In particular, one must measure the current and the gradient of the phase
in the center of the system.
The phase varies linearly in the central region of the chain, and this
allows a numerical calculation of its gradient. It can also be shown
that the effective superfluid weight can be extracted 
from the quantity \cite{Maslov 1996}:
\begin{equation}
J=\int_{0}^{L}j(x)dx=\tilde{D}(L)\phi.
\label{bigJ}
\end{equation}
In our calculations we simply replaced the integral by a sum over all the links.  
We find that the results obtained using the two approaches agree to within $1\%$.



\begin{centering}
\begin{figure}
\epsfig {file=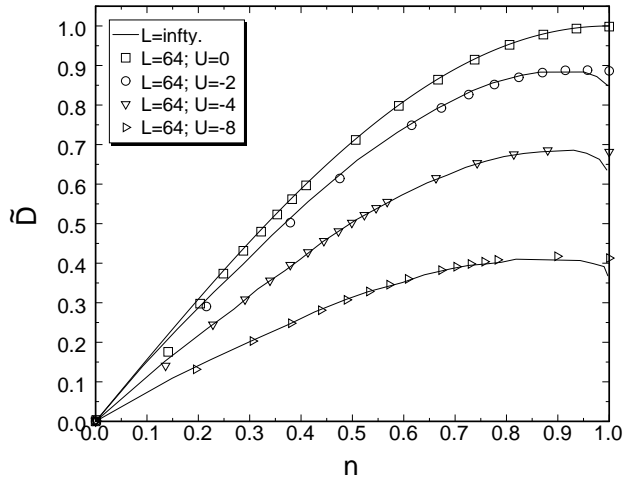,width=70mm,angle=-90}
\caption{Effective superfluid weight of a Hubbard chain ($L=64$) connected to
smooth contacts of width $\ell_c=20$, as a function
of density $n$, and for different values of $U$. We add for comparison the Bethe Ansatz results in the thermodynamic limit (solid lines).
}
\label{fig6}
\end{figure}
\end{centering}

Figure \ref{fig6} shows the results for the effective superfluid weight 
$\tilde{D}$ for a Hubbard chain of length $L=64$ with contacts of width
$\ell_c=20$. As in Fig.\ref{fig4}, the solid lines are
the Bethe-Ansatz results for $D_s$ in the thermodynamic limit.  As one can 
see, the DMRG results are in close agreement with the Bethe-ansatz 
results, except for $n=1$. It may be that logarithmic contributions 
affect the convergence of the DMRG for $n=1$\cite{Kawakami and Yang}. 
The extended
contact approach provides a much closer match between the 
supeconducting leads and the Hubbard chain so that we have 
essentially achieved perfect Andreev boundary conditions without any need 
for tuning of parameters. 

\section{Conclusions}

Here we have reported results of a numerical study of a one-dimensional 
Hubbard model coupled to external pair fields. DMRG calculations typically 
use open boundary conditions making it simple to couple the ends of an 
interacting system to a classical potential or magnetic fields. Here we 
have explored the numerics involved in coupling to a quantum pair field which 
can inject or remove pairs of electrons. We have seen how the pair 
transport varies as a function of the interaction $U$, the filling 
 $\langle n \rangle$ and the length of the Hubbard chain. Various 
current-phase relations associated with the degree of Andreev reflection 
were clearly seen. A phenomenological effective superfluid weight $\tilde{D}$ 
was introduced and found to be in close agreement with Bethe-ansatz results for 
the superfluid weight of an infinite ring.

Finally, the idea of an extended pair transfer contact was introduced. 
This was found to provide a useful way to effectively match the pair field 
injection such that the Andreev reflection approached unity. This is reminiscent
of the extended tapered connections used to match waveguides with 
different propagation characteristics and may prove useful in obtaining 
optimal matching of bulk leads to nanowires.

\acknowledgments

SRW acknowledges the support of the NSF under grant DMR03-11843, and DJS 
acknowledges support from the Center of Nanophase Material Science at 
Oak Ridge National Laboratory (Tennessee). We would like to 
thank Ian Affleck for insightful discussions.


\begin{thebibliography}{99}

\bibitem{Maslov 1996} D. L. Maslov, M. Stone, P. M. Goldbart, and D. Loss, Phys. Rev. B {\bf53}, 1548 (1996)
                                                                                
\bibitem{Fazio 1996} R. Fazio, F. W. J. Hekking, and A. A. Odintsov, Phys. Rev. B {\bf53}, 6653(1996)

\bibitem{Takane 1997} Y. Takane, J. Phys. Soc. Jpn. {\bf 66}, 537 (1997)

\bibitem{Affleck 2000} I.~Affleck, J-S.~Caux and A.M.~Zagoskin, {\sl Phys.~Rev.~B} {\bf 62},
1433 (2000).



\bibitem{Kozub 2003} V.I.~Kozub, A.V.~Lopatin, and V.M.~Vinokur, 
{\sl Phys.~Rev.~Lett.}  {\bf 90}, 226805 (2003).

\bibitem{Ishii 1970} C.~Ishii, {\sl Prog.~Theor.~Phys.} {\bf 44}, 1525 (1970).

\bibitem{Caux 2002} J.S.~Caux, H.~Saleur, and F.~Siano, 
{\sl Phys.~Rev.~Lett.} {\bf88}, 106402 (2002).

\bibitem{Kawakami and Yang} N. Kawakami and S.-K. Yang, Phys. Rev. B {\bf44}, 7844-7851 (1991).

%


(1994).

%
%
                                                                                
                                                                                
\bibitem{White 1992} S.R. White, Phys. Rev. Lett. {\bf69}, 2863 (1992)
                                                                                
\bibitem{White 1993} S.R. White, Phys. Rev. B {\bf48}, 10345 (1993). See also U. Schollw\"ock. Rev. Mod. Phys. {\bf77}, 259 (2005)

\bibitem{threeleg} S.R. White and D.J. Scalapino, Phys. Rev.  B{\bf 57}, 3031 (1998).

\bibitem{extrap} G. Hager, G. Wellein, E. Jeckelmann, and H. Fehske, Phys. Rev. B {\bf 71}, 075108 (2005)

\bibitem{Vekic 1993} M. Vekic and S.R.~White, Phys. Rev. Lett. {\bf71}, 4283 (1993).

\bibitem{smoothfun} Asymptotically, as $\ell_c\to\infty$, one would expect a 
smoothing function with all derivatives continuous to be superior to this 
function which has a discontinuity in the second derivative. However, for small
to moderate $\ell_c$ this function gives excellent results.
                                                                                
                                                                                
\end{thebibliography}
\end{document}